\documentclass[aps,
pre,
reprint,
nofootinbib,
 amsmath,amssymb,
]{revtex4-2}

\usepackage{graphicx}
\usepackage{array}
\usepackage{dcolumn}
\usepackage{bm}
\usepackage{hyperref}

\usepackage[T1]{fontenc}
\usepackage{siunitx}
\DeclareSIUnit\pixel{pixels}

\newcommand{\new}[1]{\textcolor{black}{#1}}

\begin{document}

\title{Effects of salinity on flows of dense colloidal suspensions}
\thanks{All data presented in this article are available on Zenodo repository. See DOI: \href{https://doi.org/10.5281/zenodo.10203682}{10.5281/zenodo.10203682} and \href{https://doi.org/10.5281/zenodo.10671985}{10.5281/zenodo.10671985}.}%

\author{Marc Lagoin}
\altaffiliation[Currently at: ]{Univ. Bordeaux, CNRS, LOMA, UMR 5798, F-33400, Talence, France}
\affiliation{Univ Lyon, Université Claude Bernard Lyon 1, CNRS, Institut Lumière Matière, F-69622, Villeurbanne, France}
 
\author{Rémy Fulcrand}%
\affiliation{Univ Lyon, Université Claude Bernard Lyon 1, CNRS, Institut Lumière Matière, F-69622, Villeurbanne, France}

\author{Agnès Piednoir}%
\affiliation{Univ Lyon, Université Claude Bernard Lyon 1, CNRS, Institut Lumière Matière, F-69622, Villeurbanne, France}

\author{Antoine Bérut}%
\email[Corresponding author: ]{antoine.berut@univ-lyon1.fr}
\affiliation{Univ Lyon, Université Claude Bernard Lyon 1, CNRS, Institut Lumière Matière, F-69622, Villeurbanne, France}

\date{\today}

\begin{abstract}
We experimentally study the effects of salt concentration on the flowing dynamics of dense suspensions of micrometer-sized silica particles in microfluidic drums. In pure water, the particles are fully sedimented under their own weight, but do not touch each others due to their negative surface charges, which results in a ``frictionless'' dense colloidal suspension. When the pile is inclined above a critical angle $\theta_c \sim \ang{5}$ a fast avalanche occurs, similar to what is expected for classical athermal granular media. When inclined below this angle, the pile slowly creeps until it reaches flatness. Adding ions in solution screens the repulsive forces between particles, and the flowing properties of the suspension are modified. We observe significant changes in the fast avalanche regime: a time delay appears before the onset of the avalanche and increases with the salt concentration, the whole dynamics becomes slower, and the critical angle $\theta_c$ increases from $\sim \ang{5}$ to $\sim \ang{20}$. In contrast, the slow creep regime does not seem to be heavily modified. These behaviors can be explained by considering an increase in both the initial packing fraction of the suspension $\Phi_0$, and the effective friction between the particles $\mu_p$. These observations are confirmed by confocal microscopy measurements to estimate the initial packing fraction of the suspensions, and AFM measurements to quantify the particles surface roughness and the repulsion forces, as a function of the ionic strength of the suspensions.
\end{abstract}

\maketitle


\section{Introduction}

Rheology of colloidal suspensions has been an active field for years, and it is well known that interactions (van der Waals, electrostatics, polymer brushes, etc.) between particles play an important role in the suspension behavior~\cite{Russel1991,Fagan1997,Mewis2011}. In the last decade, numerous experimental and numerical works have led to a complete theoretical framework to describe the rheology of dense non-Brownian suspensions~\cite{Guazzelli2018}. The main difference with colloidal suspensions is that bigger particles are not sensitive to thermal motion, and that surface forces are usually negligible. However, surface interactions between grains have recently been proposed as the key ingredient to explain several macroscopic rheologic behavior, such as shear-thickening~\cite{Seto2013,Wyart2014} or shear-thinning~\cite{Lemaire2023} in non-Brownian suspensions. It is therefore interesting to study how changes in particles interactions can affect the flowing properties of the suspensions. In colloidal suspensions, salt concentration has been historically used as a way to tune the interactions between particles~\cite{Yethiraj2003,Zaccarelli2007}. In non-Brownian suspensions, salt has also been used recently as a way to \new{effectively} enhance the friction between particles~\cite{Clavaud2017,Perrin2019}, \new{but the exact microscopic effect of the salt remains unclear}.

The cross-over between thermal ``colloids'' and athermal ``granular'' suspensions is controlled by the gravitational Péclet number:
\begin{equation}
\label{eq:Peclet}
P_e = \frac{mgd}{k_\mathrm{B}T}
\end{equation}
with $d$ the diameter of the particles, $m = \frac{\Delta\rho}{6}\pi d^3$ the mass the particles corrected by the buoyancy ($\Delta\rho = \rho_\mathrm{silica} - \rho_\mathrm{fluid}$ is the difference of density between the particle and its surrounding fluid), $g$ the gravitational acceleration, $k_\mathrm{B}$ the Boltzmann constant, and $T$ the temperature.
In this work, we place ourselves in the intermediate regime of ``dense colloidal suspensions'' ($P_e \gtrsim 1$), where the particles are fully sedimented, inducing a high concentration in particles, but the thermal agitation and the surface interactions cannot be neglected~\cite{Trulsson2015,Wang2015,Berut2019}. In particular, we use silica micro-particles that are negatively charged and show a repulsive interaction in water, that we can tune by adding ions in solutions~\cite{Israelachvili2011}. In a previous work~\cite{Berut2019} we have shown that such dense colloidal suspensions show peculiar flow properties when inclined in rotating drum experiments. Above a threshold angle $\theta_c$, those suspensions exhibit a ``fast avalanche'' regime, that is similar to the one observed in non-Brownian ones. Below $\theta_c$ they show a ``slow creep'' regime, that is thermally activated and depends heavily on the Péclet number. 

\new{In this study, we explore the flowing behavior of such dense colloidal suspensions when the repulsive interactions between the particles is progressively screened by salt added in the suspension. We focus in particular on the transition region between totally repulsive ``frictionless'' particles, and totally adhesive ``non-flowing'' particles. We interpret our results with the theoretical framework developed for non-Brownian suspensions, and we perform additional microscopic measurements of surface roughness, repulsion forces, and piles compacity, in order to connect the rheology of the suspensions to the microscopic interactions between the particles.}

\section{Experimental set-up}

\subsection{Microfluidic drums}

Suspensions are made of silica particles from \textit{\href{https://microparticles.de/}{microParticles GmbH}}, with diameter $d = \SI{2.12(6)}{\micro\meter}$, dispersed in solutions  of NaCl (\textit{Sigma-Aldrich}) in deionized water (\textit{ELGA Purelab\textregistered Flex}, \SI{18.2}{\mega\ohm\cm}) with concentration $C$ ranging from $0$ to \SI{2.5e-2}{\mole\per\liter}. The gravitational Péclet number of those particles in suspension is $P_e \approx 21$. The suspensions are held in polydimethylsiloxane (PDMS) microfluidic drums, made with standard soft-lithography techniques. The drums have a diameter of \SI{100}{\micro\meter} and a depth of \SI{45}{\micro\meter}.

The microfluidic drums are filled using the following protocol: a PDMS sample is made with an array of thousands circular holes with the desired diameter and depth (once sealed, these holes will become the drums which contain the colloidal suspension). The PDMS sample is carefully washed and rinsed, first with isopropyl alcohol, then with deionized water. It is then cleaned for \SI{15}{\minute} in deionized water in an ultrasonic bath. Next, it is immersed in a beaker containing the saline solution with the desired NaCl concentration $C$, and is let to degas for \SI{15}{\minute} in the ultrasonic bath. The PDMS sample is removed from the ultrasonic bath and placed on a sample holder, the drums facing up. At this stage, the drums are only filled with the saline solution, and a drop (\SI{200}{\micro\liter}) of this solution is added on top of the sample, to avoid bubble formation due to evaporation. Then a droplet (\SI{30}{\micro\liter}) of a concentrated microparticles suspension is injected with a micropipette on top of the microdrums. The particles are let to sediment for \SI{1}{\minute} in the drums. Finally, the drums are closed by placing a clean glass slide\footnote{The cleaning procedure for the glass slide is the same as the one for the PDMS sample.} on top of the PDMS sample, and by pressing it against the PDMS. The glass slide is maintained in position by six screws in the sample holder, which guarantees that the drums remains correctly sealed during the whole experiment. The particles typically fill $\sim$\SI{25}{\percent} of the drums volume.

\begin{figure}[ht!]
\centering
  \includegraphics[width=0.48\textwidth]{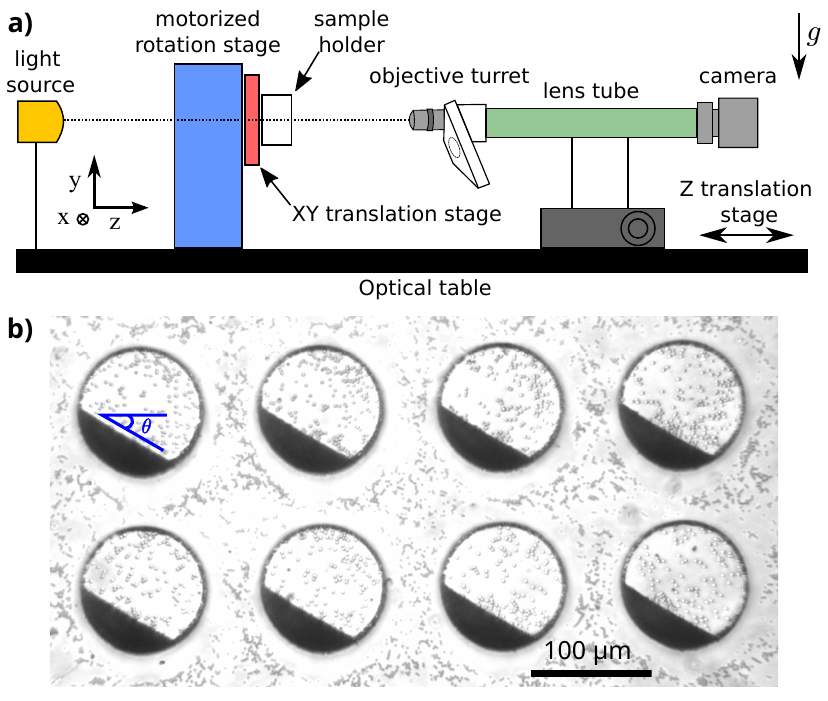}
  \caption{\textbf{a)} Schematic representation of the horizontal video-microscopy apparatus, made to visualize the flow of the dense colloidal suspensions in vertically held microfluidic drums. \textbf{b)} Example of experimental image measured during the pile relaxation experiment: an array of 8 drums filled with $\sim$\SI{25}{\percent} of colloidal suspension made of \SI{2.12}{\micro\meter} particles in \SI{1e-3}{\mole\per\liter} NaCl solution. The measured angle $\theta$ is shown in blue on the first drum. Note that the field of view has been cropped for readability, and that real images used during experiments have 20 drums in the field of view.}
  \label{fig:exp_setup}
\end{figure}

The observation is made with the custom-made experimental set-up shown in Figure~\ref{fig:exp_setup} \textbf{a)}. It is a horizontal video-microscopy apparatus, made of a CCD camera (\textit{Basler acA2440-75um}) linked to a microscope turret with long working distance microscope objectives (\textit{Olympus MPLFLN x10}, and \textit{LUCPLFLN x20}) through a lens tube (\textit{InfiniTube\texttrademark Standard}), in front of a motorized rotation stage (\textit{Newport URB100CC}), with a manual 2D translation (\textit{Owis KT 90-D56-EP}) for the sample holder. To guarantee correct visualization of the sample, the rotation axis of the rotation stage is aligned with the optical axis of the video-microscopy system with a very high precision (up to a few microns). This axis is horizontal, so that the field of view contains the vertical gravity vector. To avoid external vibration, the whole set-up in installed on an optical table with passive isolation mounts (\textit{Thorlabs PWA075}).

Before each measurement, the sample is shaken so that the particles are suspended, then let to sediment for \SI{8}{\minute}, ensuring that the initial horizontal state of the pile is the same for each experiment. Then, the drums are rotated by an initial angle $\theta_S = \SI{30}{\degree}$ and images are taken with a logarithmic framerate for \SI{24}{\hour} while the pile relaxes toward horizontal (at the beginning of the experiment the frame-rate is $20$ images per second, at the end it is $10^{-3}$ images per second). An example of experimental image is shown in Figure~\ref{fig:exp_setup} \textbf{b)}. Thanks to the use of low magnification microscope objectives, we are able to record simultaneously the flows in 20 different drums for each experiment. Images are then analyzed using contrast difference (contour finding algorithm from \textit{scikit-image}) to automatically detect the top surface of the pile, and extract its angle with respect to the horizontal $\theta$ as a function of the time $t$. \new{The typical dispersion between the 20 different angles $\theta_i(t)$ obtained for the 20 different drums in the same experiment is of a few degrees (see fine brown curves in Figure~\ref{fig:main_result}). This leads to an accuracy of the average angle better than \SI{0.5}{\degree}.}

\subsection{AFM Measurements}

AFM force spectroscopy studies are performed on a MFP-3D (from \textit{Asylum Research, Oxford Instrument}) with a homemade colloidal probe~\cite{Audry2010}. The probe is a silica bead of \SI{10}{\micro\meter}, from the same manufacturer (\textit{\href{https://microparticles.de/}{microParticles GmbH}}), glued to the end of a silicon nitride cantilever (\textit{DNP cantilever - Bruker}) with a bi-component adhesive (\textit{Loctite EA3430}). We used the thermal noise method~\cite{Hutter1993,Butt2005} to know the nominal spring constant of the tooled cantilever for quantitative measurements.

The force curves are recorded in deionized water and in three different NaCl solutions: forces between the silica probe and a flat silica substrate are measured during the movement of the probe at constant velocity of \SI{1}{\micro\meter\per\second} towards the surface until contact is made (the maximum applied force is \SI{6}{\nano\newton}) and then during the return.

Surface imaging of \SI{2}{\micro\meter} silica particles is also performed in tapping mode (PPP NCHR AFM tip from \textit{NanoAndMore}). Typical images have a spatial resolution of \SI{2}{\nano\meter\per\pixel}, and the total field of view is about $1\times1$~\si{\square\micro\meter} ($512\times512$ pixels). To determine their RMS roughness value, the curvature of the spherical cap of the bead was subtracted from the image by a flatten of order 2 along the X and Y axis.

\subsection{Confocal Microscopy}

\begin{figure}[ht!]
\centering
  \includegraphics[width=0.4\textwidth]{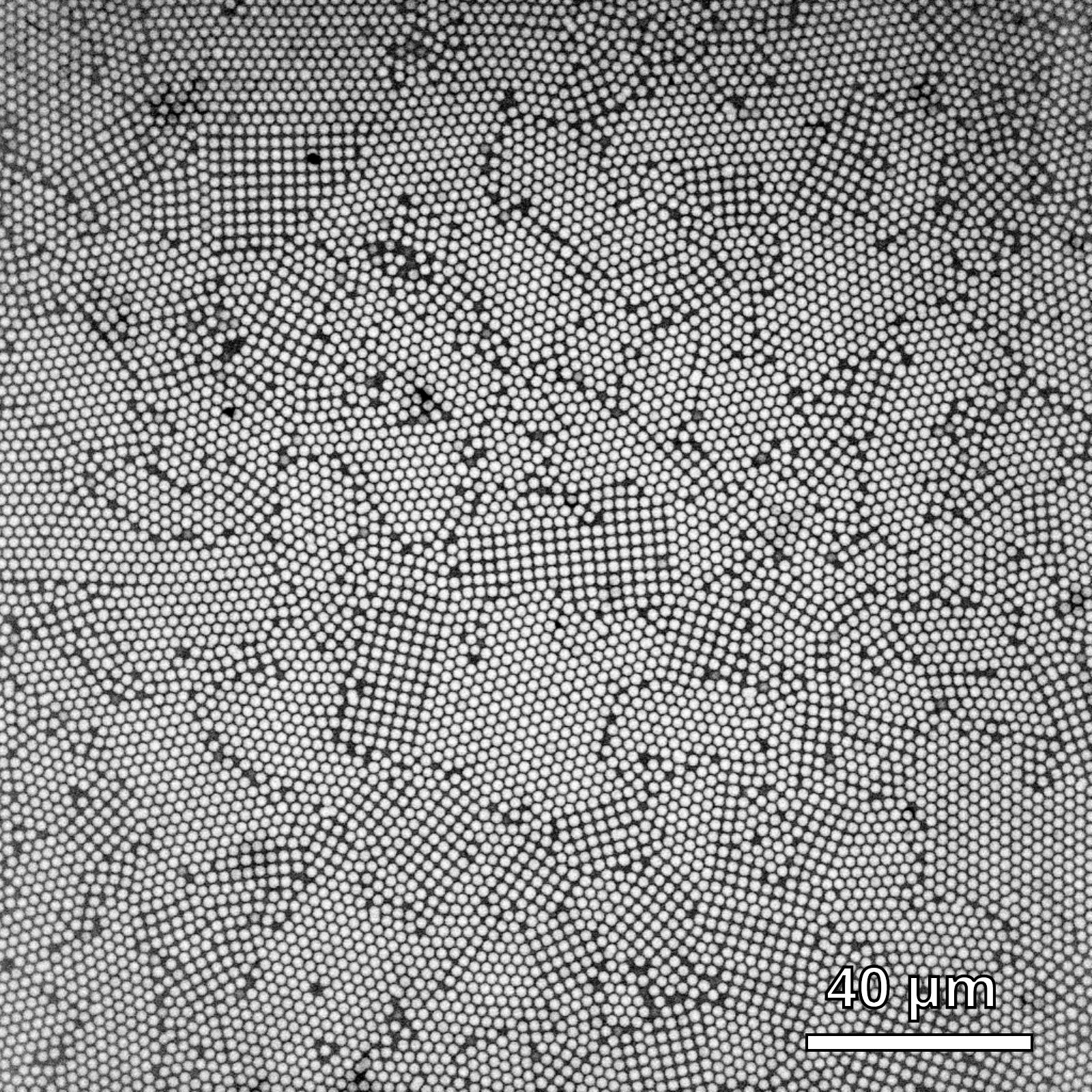}
  \caption{Image of the bottom layer of the suspension of \SI{2.12}{\micro\meter} particles in deionized water, observed with the confocal microscope. The contrast has been inverted so that the particles appear light on a dark background. Poly-crystalline structure is visible, due to the relatively high P\'{e}clet number and high monodisperisty of the particles.}
  \label{fig:confocal}
\end{figure}

Stacks of images of dense colloidal suspensions at rest are obtained with a confocal microscope (\textit{Leica SP5}, excitation wavelength \SI{488}{\nano\meter}). The measurement is made with two different salt concentration (``pure'' deionized water and \SI{1e-2}{\mole\per\liter} NaCl). Each suspension is introduced in a container, and a small amount of Rhodamine B is added (concentration \SI{6e-6}{\mole\per\liter}). Before measurement, the suspension is let to sediment for \SI{10}{\minute}. Scanns are performed at \SI{400}{\hertz}, with an oil-immersion microscope objective (\textit{HCX PL FLUOTAR  63.0x 1.25 OIL}). The images stacks have a spatial resolution of \SI{0.2}{\micro\meter\per\pixel} in all three directions (XYZ), and the total field of view is about $200\times200\times10$~\si{\cubic\micro\meter}. Before tracking the particles' coordinates, the contrast of each image is corrected. An example of the image obtained is shown in Figure~\ref{fig:confocal}. Due to the relatively high monodisperisty of the particles, and the high P\'{e}clet number, the sediment shows a poly-crystalline structure, as it has been observed in other experimental~\cite{Hoogenboom2003,Nakamura2021} or numerical~\cite{Marechal2011} works.

We use the TrackPy software~\cite{trackpy} to obtain the 3D coordinates of the particles present in the stack. After analysis, about 40000 particles distributed on up to 5 successive layers of sediment are found.

\section{Results}

\subsection{Microfluidic drums}

\begin{figure}[ht!]
\centering
  \includegraphics[width=0.48\textwidth]{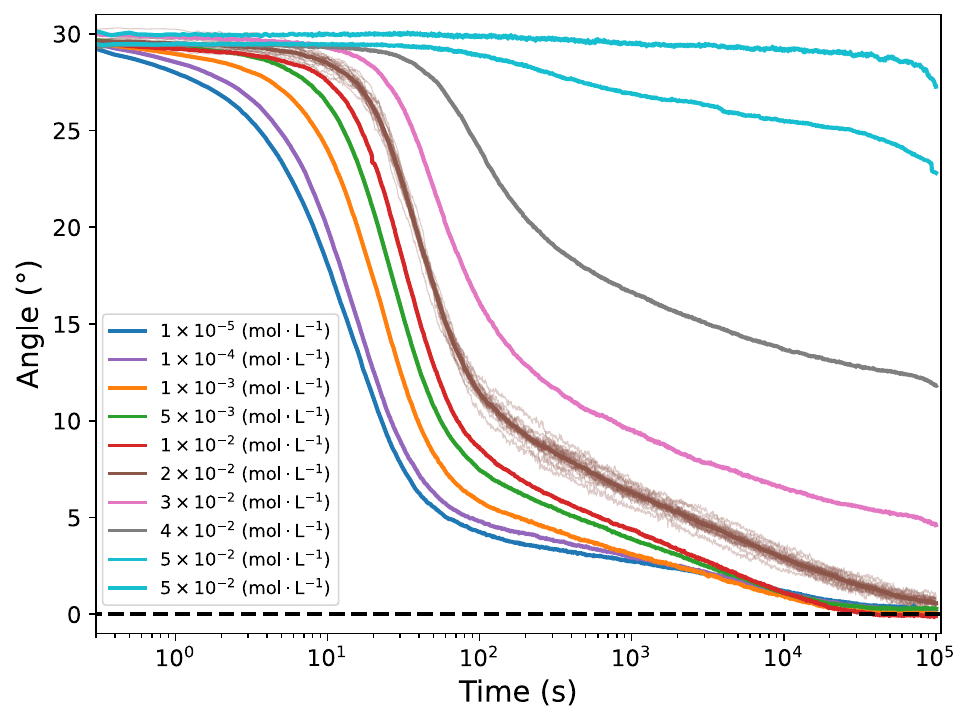}
  \caption{Average flow curves for dense suspensions of \SI{2.12}{\micro\meter} silica particles inclined with same starting angle $\theta_S = \SI{30}{\degree}$, in salty solutions with different concentrations. Values of the NaCl concentration is indicated in legend: the bottom curve (dark blue) corresponds to the solution with the lowest concentration (deionized water, ionic strength estimated $\sim\SI{1e-5}{\mole\per\liter}$), and the top curve (light blue) corresponds to the solution with the highest concentration (ionic strength $= \SI{5e-2}{\mole\per\liter}$). Each curve is obtained by averaging pile angles $\theta(t)$ measured on the 20 drums visible in a single experiment.\new{To help visualize the typical data dispersion, the 20 individual flow curves used to compute the average are showed for the NaCl concentration $\SI{1e-2}{\mole\per\liter}$ (fine brown curves).}}
  \label{fig:main_result}
\end{figure}

The effect of the salt concentration on the average flow curves of dense colloidal suspensions in micro-fluidic drums is presented in Figure~\ref{fig:main_result}. For the lowest ionic strength\footnote{For a monovalent electrolyte such as NaCl, the ionic strength is directly equal to the concentration.} (bottom curve), the typical flow behavior is retrieved, with a ``fast avalanche'' regime at high angle, and a ``slow creep'' regime, which is logarithmic in time, below a threshold angle $\theta_c$~\cite{Berut2019}. However, when the ionic strength is increased, the flow behavior progressively changes, as the repulsive force between the particles is progressively screened. The most noticeable changes are: the increase of a waiting time plateau at the beginning of the experiment before the fast avalanche regime starts, the increase of the time needed for the fast avalanche regime to reach the threshold angle $\theta_c$, the increase of the threshold angle $\theta_c$ itself, and the increase of the final angle at the end of the experiment. There might also be a small change in the slow creep regime, however, as discussed later, this effect is less clear. For the highest ionic strength of \SI{5e-2}{\mole\per\liter}, almost no flow is observed when the drums are rotated\footnote{Note that the steep decrease that is visible at the very end of the high ionic strength curves (at $t \approx \SI{1e5}{\second}$) is an artifact due to the aging of the micro-fluidic sample.}, and particles agglomerates are visible in the pile. This is consistent with the fact that ionic strengths above \SI{5e-2}{\mole\per\liter} can be used to generate floculated suspensions~\cite{Fusier2018} or colloidal gels~\cite{Manley2005} from silica particles dispersed in water. Those agglomerates of particles also lead to irreproducible avalanche curves (as shown in Figure~\ref{fig:main_result}), because they can sustain non-flat pile interfaces where the angle $\theta$ is not properly defined anymore.

\begin{figure}[ht!]
\centering
  \includegraphics[width=0.48\textwidth]{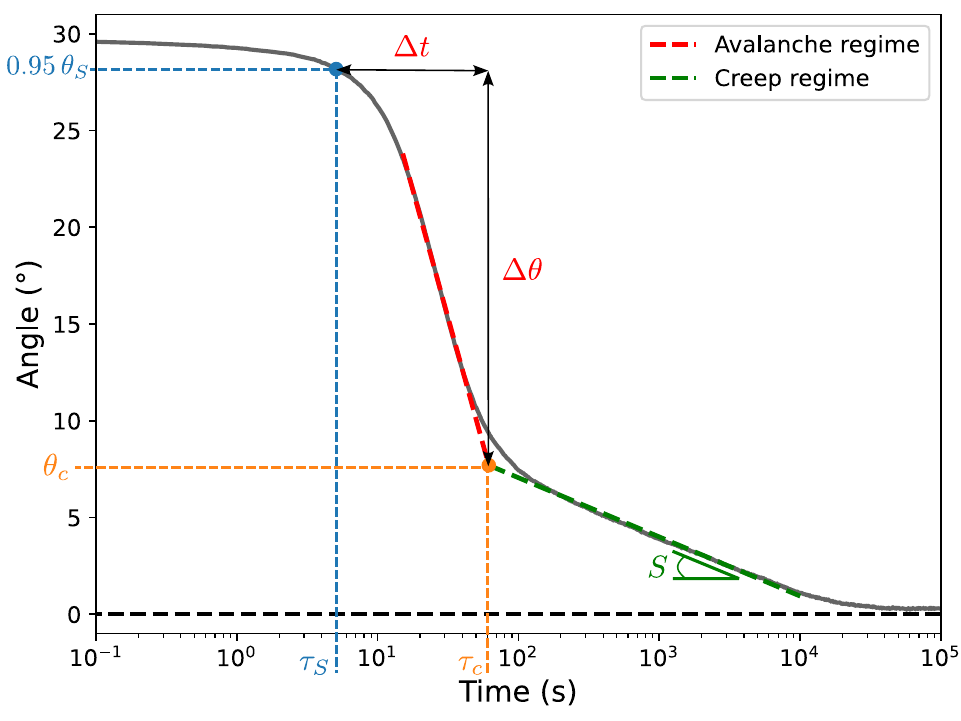}
  \caption{Schematical description of the quantities of interest measured for each flow curve. $\theta_S$ is the initial starting angle ; $\tau_S$ is the starting time needed to reach $0.95 \, \theta_S$ ; $\Delta \theta / \Delta t$ is the average speed of the avalanche ; $\theta_c$ is the threshold angle between the two regimes ; and $S$ is the slope of the creep regime. }
  \label{fig:def_quantities}
\end{figure}

To better quantify the effect of the salt concentration, we define a few experimental quantities, schematically presented in Figure~\ref{fig:def_quantities}. We call $\tau_s$ the ``starting time'' of the fast avalanche, that is the time required for the pile to reach \SI{95}{\percent} of its initial angle $\theta_S$. We fit both the end of the fast avalanche regime, and the slow creep regime by a linear function in the semilogarithmic plot (\textit{i.e.} $\theta = A \log(t) + B$ with $A$ and $B$ two constants). We define the threshold angle $\theta_c$ as the crossing point between the two fitted regimes. We call ``avalanche speed'' $\Delta\theta/\Delta t$, the average flowing rate of the fast avalanche regime. We call $S$ the slope of the slow creep regime in the semilogarithmic time-scale. \new{Note that those four quantities are defined in different temporal regions of the flow curve, and are mathematically independent one from another.}

\begin{figure}[ht!]
\centering
  \includegraphics[width=0.48\textwidth]{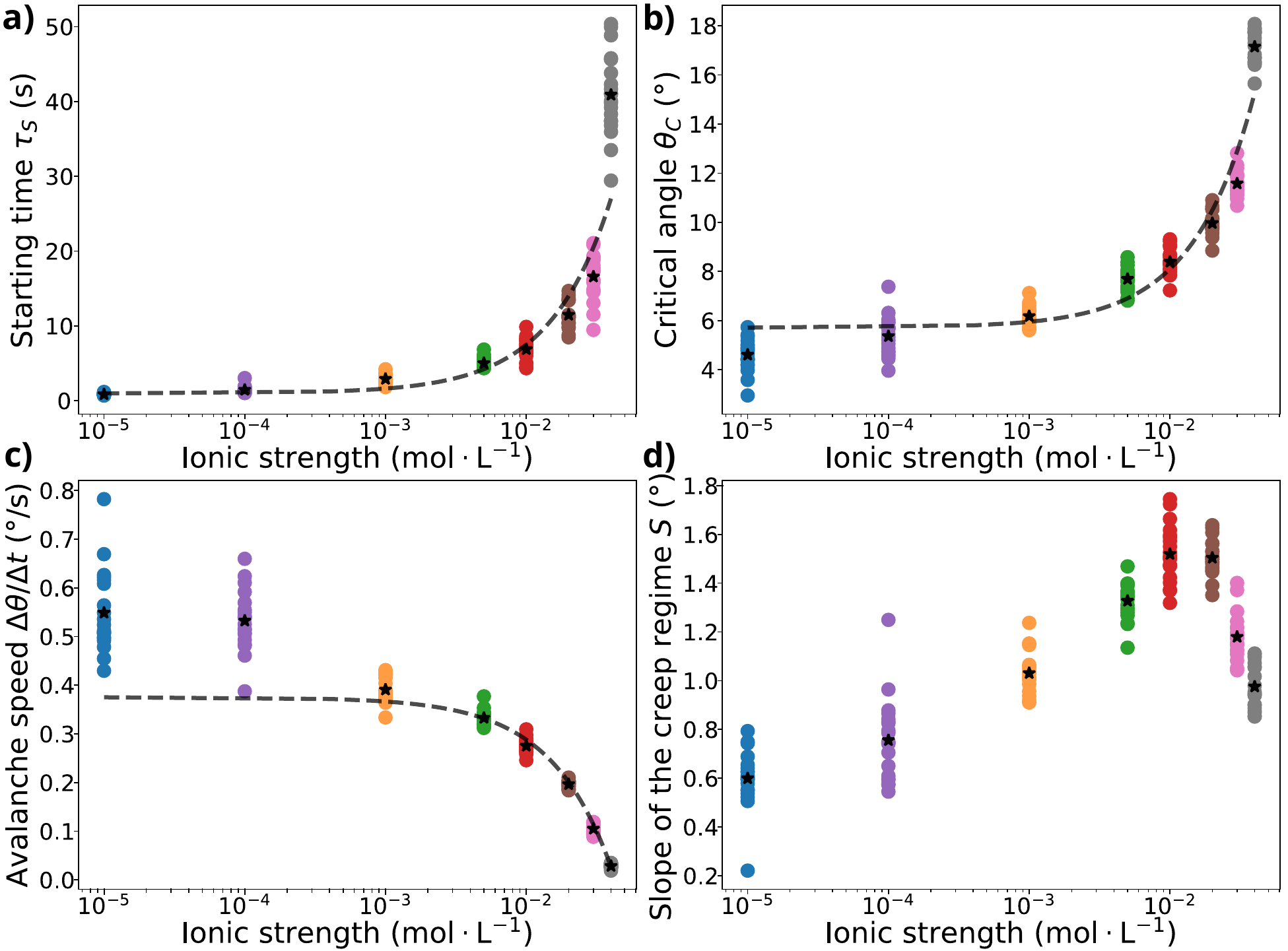}
  \caption{Measured values of a) the starting time $\tau_S$, b) the critical angle $\theta_c$, c) the avalanche speed $\Delta \theta / \Delta t$, and d) the slope of the creep regime in semilogarithmic time scale $S$, as a function of the ionic strength of the suspension. Color points are individual measurements, black stars are the averaged values. \new{Gray dashed lines corresponds to a linear fit where each average point is weighted by the inverse of its standard deviation.}}
  \label{fig:mes_quantities}
\end{figure}

The measured values are presented in Figure~\ref{fig:mes_quantities}. Both $\tau_S$ and $\theta_c$ increase \new{slightly faster than linearly} with the ionic strength, with a steep increase when the ionic strength is close to \SI{5e-2}{\mole\per\liter}. The avalanche speed $\Delta\theta/\Delta t$ seems to decrease almost linearly with the ionic strength. Finally, the slope of the creep regime $S$ seems to first increase up to a maximum when the ionic strength is close to \SI{1e-2}{\mole\per\liter}, and then slightly decreases.

\subsection{AFM Measurements}

\begin{figure}[ht!]
\centering
  \includegraphics[width=0.48\textwidth]{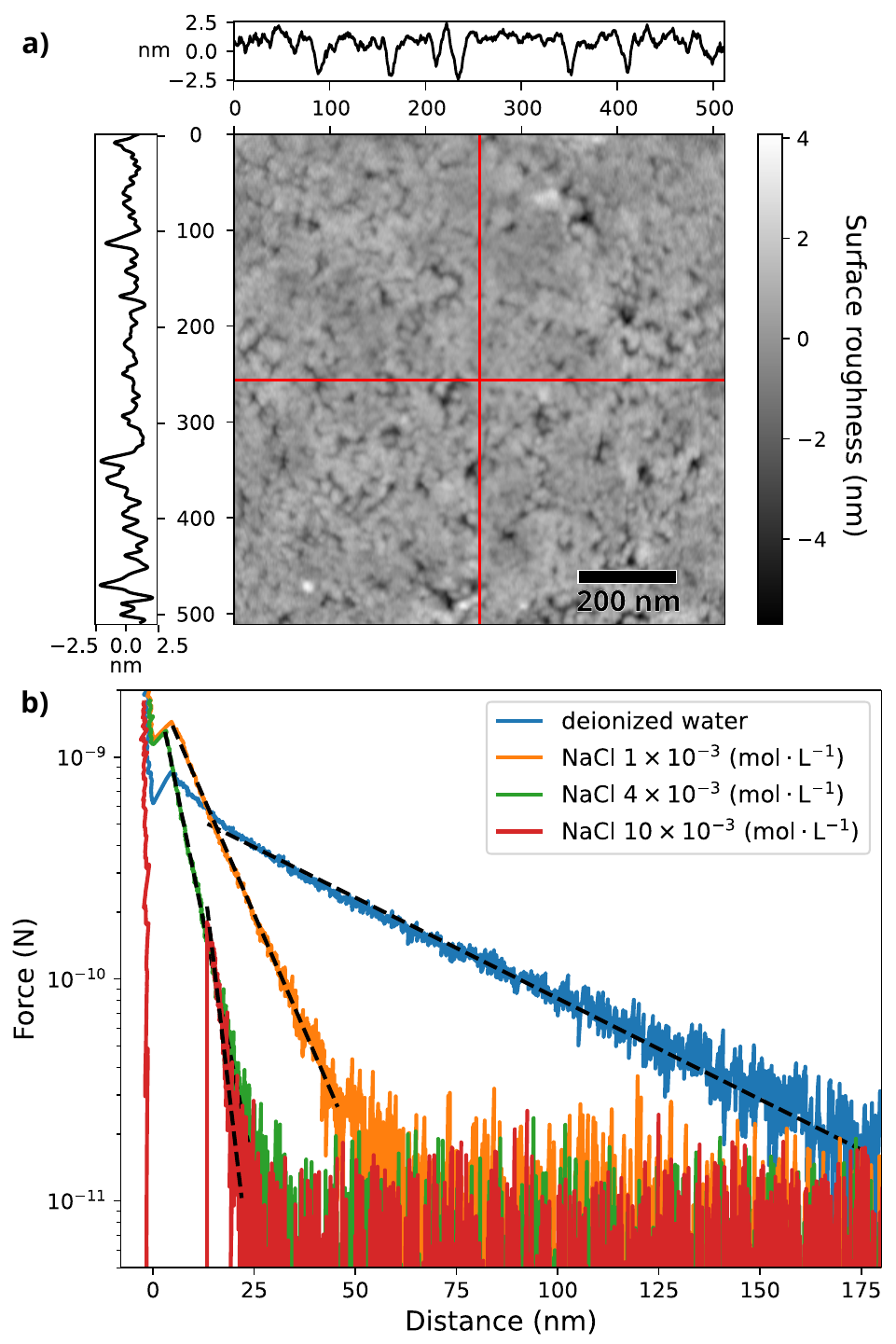}
  \caption{\textbf{a)} AFM imaging of a \SI{2}{\micro\meter} silica particle's surface roughness (the lateral resolution is \SI{2}{\nano\meter\per\pixel}, and total field of view is $1\times1$~\si{\square\micro\meter}). The topographic profiles extracted from the image along a vertical and horizontal lines are respectively shown to the left and on top of the image. \textbf{b)} Repulsive forces measured between a \SI{10}{\micro\meter} silica particle and a flat silica surface, for different concentrations of the saline solution. The black dashed-lines correspond to numerical fits with the theoretical formula for the electrostatic double-layer interaction force between a sphere and a flat surface (see eq. ~\ref{eq:EDL} and table~\ref{tab:AFM} for best fitting parameters values).}
  \label{fig:AFM}
\end{figure}

The colloids surface roughness is measured by AFM imaging. The RMS roughness of a \SI{2}{\micro\meter} silica particle is less than \SI{1}{\nano\meter} over $\SI{1}{\micro\meter} \times \SI{1}{\micro\meter}$. A typical image of the surface imaging is shown in Figure~\ref{fig:AFM}~\textbf{a)}.

\new{Typical} repulsive forces $F$ are shown in Figure~\ref{fig:AFM}~\textbf{b)} as a function of the distance $D$ between the particle's surface and the flat silica surface, for different salt concentrations. \new{At large enough separation distance,} they show a good agreement with the theoretical electrostatic double-layer forces between surfaces in liquids \new{at constant potential}~\cite{Israelachvili2011}:
\begin{equation}
\label{eq:EDL}
F(D) = \frac{d}{2\lambda_\mathrm{D}} \, Z \, \mathrm{e}^{\frac{-D}{\lambda_\mathrm{D}}}
\end{equation}
where, $d$ is the particle diameter, $\lambda_\mathrm{D}$ is the Debye length, and $Z$ is a constant equal to $9.22 \times 10^{-11} \tanh^2(\psi_0/103)$ \si{\joule\per\meter} at \SI{25}{\celsius}, with $\psi_0$ the surface potential in \si{\milli\volt}.

\new{Dashed lines in Figure~\ref{fig:AFM}~\textbf{b)} are the numerical fits of the forces curves $F(D)$ with eq.~\ref{eq:EDL}, using two free parameters ($\psi_0$ and $\lambda_\mathrm{D}$). The average best fitting values obtained for the surface potential $\psi_0$ and the Debye length $\lambda_\mathrm{D}$ are presented in table~\ref{tab:AFM}. For each salt concentration, at least 4 independent force curves have been measured. The errorbars are estimated from the fits accuracy and data dispersion. Values of the surface potential $\psi_0$ are expected to be found between \SI{-70}{\milli\volt} and \SI{-20}{\milli\volt} in pure water~\cite{Horn1989,Ducker1992,Yamamoto2010,Luderitz2013}, and to decrease with the salt concentration~\cite{Horn1989,Ducker1992}. Note that the exact value depends on the surface state (cleanness, roughness) of the silica~\cite{Yamamoto2010}, and that theoretical values can be hard to determine~\cite{Hartkamp2018}. On the contrary, the expected values of the Debye lengths can easily be computed from the ionic strength of the solutions~\cite{Israelachvili2011}: at \SI{25}{\celsius}, $\lambda_\mathrm{D} = 0.304/\sqrt{C}$ \si{\nano\meter}, with $C$ the monovalent salt concentration in \si{\mole\per\liter}. The expected values are presented in table~\ref{tab:AFM} and show a good agreement with the ones measured experimentally. Note that the ionic strength of the suspension made with ``pure'' deionized water is unknown, but is expected to be about a few \SI{e-5}{\mole\per\liter} due to water contamination from dissolved carbon dioxyde~\cite{Ducker1992} and from the colloids themselves. The value $\lambda_D \approx \SI{65}{\nano\meter}$ that we find corresponds to a ionic strength of $\SI{2e-5}{\mol\per\liter}$.}

\begin{table}[ht!]
  \caption{\new{Average} best fitting parameters for surface potential $\psi_0$ and Debye length $\lambda_D$ obtained from AFM repulsive force measurements \new{(errorbars are obtained from the fit accuracy and data dispersion)}.}
  \label{tab:AFM}
  \begin{tabular}{|c|c|c|c|}
    \hline
    $C$ & $\psi_0$ & fitted $\lambda_\mathrm{D}$ & predicted $\lambda_\mathrm{D}$ \\
    (\si{\mole\per\liter}) & (\si{\milli\volt}) & (\si{\nano\meter}) & (\si{\nano\meter}) \\
    \hline
    deionized water & \new{$-27 \pm 1$} & \new{$65.4 \pm 11.2$} & n/a \\
    $1 \times 10^{-3}$ & \new{$-24 \pm 1$} & \new{$10.4 \pm 1.0$}  & 9.6 \\
    $4 \times 10^{-3}$ & \new{$-16 \pm 1$} & \new{$4.75 \pm 0.84$} & 4.81 \\
    $10 \times 10^{-3}$ & \new{$-36 \pm 20$}  & \new{$3.01 \pm 1.01$} & 3.04 \\
    \hline
  \end{tabular}
\end{table}

\subsection{Confocal Microscopy}

\begin{figure}[ht!]
\centering
  \includegraphics[width=0.48\textwidth]{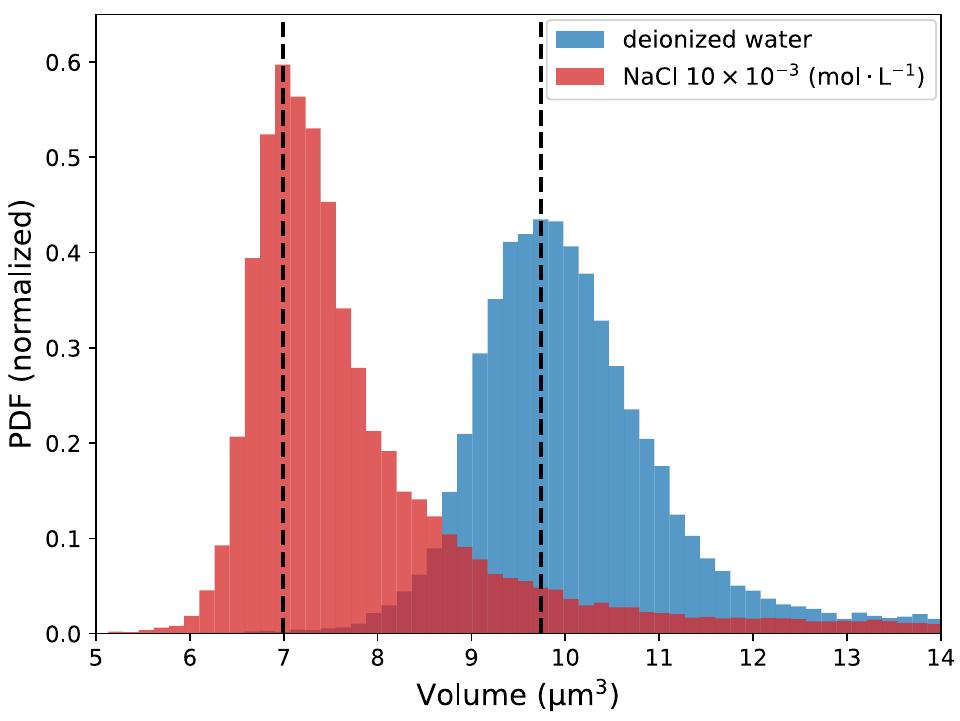}
  \caption{Distribution of the volume occupied by each particle computed from the 3D Voronoï tesselation of particles 3D coordinates obtained by confocal microscopy, for two different salt concentrations.\new{Black dashed-lines indicate the mode of each distribution.}}
  \label{fig:volume_PDF}
\end{figure}

To estimate the typical volume that is occupied by one particle on the bulk of each suspension, we compute the 3D Voronoï tessellation (\textit{scipy.spatial} library based on the \textit{Qhull} library) of the particles 3D coordinates obtained by confocal microscopy. We first remove the particles on the side/edges of the sample, then compute the distribution of volumes of the Voronoï cells of the remaining particles. The probability density functions (PDF) of the volume occupied by a single particle in the bulk are shown in figure~\ref{fig:volume_PDF}, for two different salt concentration.

Finally, the mode of the distribution is taken as the typical volume around each particle. This value is used to compute the initial packing fraction $\Phi_0$ of the suspension, given that the actual volume of one particle is know ($\pi d^3/6$). The measured packing fractions are presented in table~\ref{tab:packing_fraction} for the two tested salt concentrations.

\begin{table}[ht!]
  \caption{Measured packing fraction of the sedimented pile of colloids.}
  \label{tab:packing_fraction}
  \begin{tabular}{|c|c|}
    \hline
   $C$ (\si{\mole\per\liter}) & $\Phi_0$ (\si{\percent}) \\
    \hline
    deionized water & 51.2 \\
    $10 \times 10^{-3}$ & 61.4 \\
    \hline
  \end{tabular}
\end{table}

\section{Discussion}

In this section, we provide physical explanations for the changes in flowing behaviors of the dense colloidal suspensions when the salt concentration is increased.

\subsection{Critical angle}

Salt added to screen repulsive interactions between silica particles has been used as a way to modify the friction between the grains in non-Brownian suspensions\cite{Clavaud2017,Perrin2019}. The main idea is that the double-layer repulsive force, which size is typically given by the Debye length $\lambda_\mathrm{D}$, prevents the particles from having frictional contacts as long as $\lambda_\mathrm{D}$ is bigger than the surface roughness of the grains $r$. When salt is added, the Debye length becomes smaller, up to a point where the particles can touch each others. This induces a frictionless to frictional transition for the suspension around $\lambda_\mathrm{D} \approx r$, which leads to shear-thickening behavior~\cite{Clavaud2017} or hysteretic flows~\cite{Perrin2019}.

This idea can be used to simply explain the observed increase of the critical angle $\theta_c$ when the ionic strength increases (Figure~\ref{fig:mes_quantities}~\textbf{b)}). Indeed, $\theta_c$ corresponds roughly to the ``angle of repose'' of the granular suspension: it's the angle below which no flow should be observed if the pile was non-Brownian. Both numerical~\cite{Zhou2001,Zhou2002, Suiker2004} and experimental~\cite{Pohlman2006} studies have shown that the angle of repose of a dry granular pile increases when the microscopic friction coefficient between the grains $\mu_p$ is increased. Therefore, one can expect that the increase in salt concentration, increases the effective friction between the particles, which then increases the angle of repose of the pile. Notably, we see that the measured critical angle $\theta_c$ in deionized water is about \SI{4.6}{\degree}, which is close\footnote{Note that the critical angle $\theta_c$, as defined in Figure~\ref{fig:def_quantities}, is always lower than the real pile angle at which the transition from the ``fast avalanche'' regime to the ``slow creep'' regime occurs.} to the angle of repose \SI{5.76}{\degree} that is observed in numerical simulation for frictionless particles~\cite{Peyneau2008}. Moreover, $\theta_c$ in the solution with the highest salt concentration (\SI{4e-2}{\mole\per\liter}) is about \SI{17.2}{\degree}, which is not too far from the repose angle of \SI{25}{\degree} for macroscopic glass beads~\cite{Makse1998}.

Our AFM measurements support this hypothesis. As shown in Fig.~\ref{fig:AFM}, the typical peak surface roughness $r$ of our particles is about \SI{2}{\nano\meter}, and the repulsive force $F(D)$ between one particle and a flat silica surface is well described by the theoretical double-layer electrostatic theory (Equation~\ref{eq:EDL}). Therefore one can estimate that the critical salt concentration where the Debye length becomes equal to the surface roughness ($\lambda_\mathrm{D} \approx r$) is about $C_c = \SI{2.3e-2}{\mole\per\liter}$. This is consistent with the critical ionic strength at which we see a transition from colloidal piles completely flowing back to horizontal $\left( C \leq \SI{1e-2}{\mole\per\liter} \right)$, to completely arrested colloidal piles $\left( C \geq \SI{5e-2}{\mole\per\liter} \right)$ (see Figure~\ref{fig:main_result}). Note that we cannot directly measure the microscopic friction coefficient $\mu_p$ between particles with our experimental set-up. However, values gathered in the literature can be found in Lee \textit{et al.}~\cite{Lee2020}: silica microparticles have a typical friction coefficient $0.03 \leq \mu_p \leq 0.1$ in milli-Q water, $\mu_p \approx 0.3$ in NaCl solution with concentration $C = \SI{1e-3}{\mole\per\liter}$, and $\mu_p \approx 0.9$ in alkaline solution with ionic strength \SI{16e-2}{\mole\per\liter}.

\subsection{Starting time}

The increase of the delay before the start of the ``fast avalanche'' regime (Figure~\ref{fig:mes_quantities}~\textbf{a)}) is reminiscent of the dilatancy effects that are observed in macroscopic granular suspensions~\cite{Iverson2005,Pailha2008,Guazzelli2018}. When a pile of grains is fully immersed in a Newtonian fluid, its flowing properties strongly depend on its initial packing fraction. After a sudden inclination, loosely packed piles tend to flow almost immediately, while densely packed piles show a time delay before the initiation of the flow. This phenomenon is explained by a pore pressure feedback scenario: a densely packed pile has to dilate before being able to flow, and during this dilation, the surrounding fluid is sucked into the granular layer, which tends to stabilize the pile and delay the start of the flow~\cite{Pailha2008}.

Since the ionic strength reduces the double-layer repulsive force between the particles, one can expect that it reduces the mean distance between particles, hence increasing the initial packing fraction $\Phi_0$ of the pile. Therefore, we can expect that the ionic stress increases the time delay $\tau_S$ before the start of the flow, due to dilatancy effects. Our set-up does not allow us to measure the packing fraction $\Phi$ of the pile during the flow, to directly observe dilatancy effects. However, our confocal microscopy measurements support the fact that the initial packing fraction $\Phi_0$ of the sedimented pile increases with the ionic strength of the suspension. As shown in table~\ref{tab:packing_fraction}, the packing fraction is about \SI{51}{\percent} in deionized water and increases to about \SI{61}{\percent} in a solution with NaCl concentration \SI{1e-2}{\mole\per\liter}. Notably, the critical packing fraction $\Phi_{0C}$ above which dilatancy effect are observed in macroscopic granular suspensions~\cite{Pailha2008} is about \SI{58}{\degree}. This is consistent with the fact that we observe almost immediate flow in deionized water ($\tau_S = \SI{0.87}{\second}$), while we observe significant start delay with high ionic strength suspensions ($\tau_S = \SI{40.9}{\second}$ for $C = \SI{4e-2}{\mole\per\liter}$).

Nevertheless, two points must be noted. First, the increase of the initial packing fraction $\Phi_0$ that is observed with the increase of the salt concentration might seem surprising. Indeed, for macroscopic granular materials, it is known that the packing fraction obtained after sedimentation of the granular medium (random loose packing) decreases when the friction between the particles increases~\cite{Shokef2003,Jerkins2008,Silbert2010}. Since we have already shown that the effective friction between the particles $\mu_p$ increases with the salt concentration, one could expect that the packing fraction would rather be lower when the ionic strength of the suspension is higher. The solution to this apparent contradiction comes from the fact that our suspensions are Brownian: we believe that the thermal agitation helps the suspension to always reach the highest accessible packing fraction (random close packing). Second, the fact that we observe dilantancy effects is itself a proof that the friction between the grains increases with the salt concentration. Indeed, numerical simulations have shown that frictionless grains do not show dilatancy effects~\cite{Peyneau2008}.

\subsection{Avalanche speed}

The fact that the salt concentration increases the effective friction between the particles can also explain the observed decrease of the avalanche speed $\Delta\theta/\Delta t$ (Figure~\ref{fig:mes_quantities}~\textbf{c)}). Indeed, both numerical simulations~\cite{Gallier2014,Mari2014,Trulsson2017,Chevremont2019,More2020} and experimental works~\cite{Tanner2016, Hoyle2020} have shown that the rheology of dense non-Brownian suspensions depends on the microscopic friction coefficient between the grains. A recent review can be found in Lemaire \textit{et al.}~\cite{Lemaire2023}, and we only recall here a few key results. For example, in volume-imposed simulations, the viscosity of the suspension $\eta_S$ increases with the microscopic friction coefficient $\mu_p$. In pressure-imposed simulations, the stress ratio $\mu$ (which can be seen as the \emph{macroscopic} friction coefficient) increases with $\mu_p$, while the volume fraction $\Phi$ decreases with $\mu_p$, at fixed viscous number $J$\footnote{For a complete definition of the viscous number $J$ used in the $\mu(J)$ rheology of dense granular suspensions, see the review by Guazzelli \textit{et al.}~\cite{Guazzelli2018}}. 

In general, it is expected that the flow rate of the suspension $Q$ decreases when the viscosity $\eta_S$ increases, when the stress ratio $\mu$ increases, and when the volume fraction $\Phi$ decreases. Thus, an increase of the microscopic friction coefficient is expected to lead to a decrease of the avalanche speed $\Delta\theta/\Delta t$. Direct comparison are difficult to achieve, since it is non-trivial to compute the theoretical flow rate $Q$ in the rotating drum geometry\footnote{Note that it is possible to predict $Q$ in simpler geometries: for example, on an inclined plane (which corresponds to pressure-imposed conditions, with constant stress ratio $\mu$), predictions yields $Q\propto J \Phi \cos\theta$, where $J$ is the viscous number and $\theta$ is the inclination angle~\cite{Billon2021}.}. But the orders of magnitude are reasonable. In our experiment we observe that $\Delta\theta/\Delta t$ decreases by a factor of $\sim 10$ between pure water ($\Delta\theta/\Delta t = \SI{0.55}{\degree\per\second}$) and high ionic strength suspensions ($\Delta\theta/\Delta t = \SI{0.03}{\degree\per\second}$ for $C=\SI{4e-2}{\mole\per\liter}$). In simulations~\cite{Chevremont2019}, the viscosity of the suspension increases by a factor of $10$ when the microscopic friction coefficient $\mu_p$ increases from $0$ to $1$ at volume fraction $\Phi=\SI{55}{\percent}$.

\subsection{Slope of the creep regime}

Following previous work~\cite{Jaeger1989,Berut2019}, the time evolution of the pile angle $\theta$ during the creep regime can be described with a simple model where particles in the top layers are considered blocked by their neighbors, and the creep occurs when they jump above those neighbors thanks to thermal agitation. This model gives the following mathematical expression:
\begin{equation}
\label{eq:creep}
\theta (t) = \frac{2}{\alpha P_e} \mathrm{arcoth} \left[ \exp \left( \frac{t}{\tau} \alpha P_e \, \mathrm{e}^{-\alpha P_e \theta_c} \right) \mathrm{coth} \left( \alpha P_e \theta_S /2 \right) \right]
\end{equation}
where: $\alpha$ is a dimensionless geometric parameter, $P_e$ is the gravitationnal Péclet number, $\tau$ is a characteristic time depending on the fluid's properties and drum geometry, $\theta_c$ is the critical angle, $\theta_S$ is the initial inclination angle (if $\theta_S \leq \theta_c$).

If $P_e \gg 1 $, and $\theta_c \ll \theta \ll 0$, the equation~\ref{eq:creep} can be approximated by:
\begin{equation}
\label{eq:creep_simple}
\theta (t) \approx \theta_S - \frac{1}{\alpha P_e} \ln \left[ 1 + \frac{t}{2\tau} \alpha P_e \, \mathrm{e}^{-\alpha P_e (\theta_c - \theta_S)} \right]
\end{equation}
Equation~\ref{eq:creep_simple} directly gives the slope of the creep regime: $S~=~1/(\alpha~P_e)$. Knowing that $P_e \approx 21.4$ for the \SI{2.12}{\micro\meter} particles and that $\alpha \approx 2.6$ was found in previous experiments~\cite{Berut2019}, we can expect $S \approx \SI{0.0177}{\radian} = \SI{1.01}{\degree}$. This is consistent with the values we measured ($\SI{0.6}{\degree} \leq S \leq \SI{1.6}{\degree}$, see Figure~\ref{fig:mes_quantities}~\textbf{d)}). 

However, the model does not predict a significant variation of $S$ with the salt concentration $C$. Indeed, when salt is added to the suspension, $P_e$ only slightly varies because the density $\rho_\mathrm{fluid}$ of the salted water varies. Even with the most concentrated solution ($C = \SI{5e-2}{\mole\per\liter}$) the density only increases by $\sim \SI{3}{\percent}$, which leads to a small decrease $P_e \approx 20.7$. As for $\alpha$, it corresponds to the ``height'' of the barrier that one particle has to cross to jump over its neighbors. One can assume that this value slightly decreases when the Debye length $\lambda_\mathrm{D}$ decreases because the particles has to jump above a particle of effective diameter $d + 2 \lambda_\mathrm{D}$. For example, if $\lambda_\mathrm{D}$ goes from \SI{50}{\nano\meter} (deionized water) to \SI{1}{\nano\meter} (high salt concentration), this would predict that $\alpha$ decreases by $\frac{0.1}{2.12} \approx \SI{5}{\percent}$. 

In the end, following the model, the slope of the creep regime $S$ should monotonically increase when the salt concentration increases, and should not vary by more than $\sim \SI{10}{\percent}$. Therefore, it remains unclear whether the variations that we observe in Figure~\ref{fig:mes_quantities}~\textbf{d)} are real physical effects, or experimental artifacts due to the difficulty to measure small pile angles during long times~\footnote{\new{Note that the decrease of $S$ that we observe for salt concentration $C \geq \SI{1e-2}{\mole\per\liter}$ (see Figure~\ref{fig:mes_quantities}~\textbf{d)}) might also come from the progressive apparition of adhesion between the grains, which is not taken into account in the simple model (Eq.~\ref{eq:creep}).}}. \new{We believe that accurate measurement of $S$ should be obtained from flow curves with an initial inclination angle $\theta_S$ below the threshold angle $\theta_C$, to avoid any influence of the transition between the creep regime and the preceding fast avalanche regime.}

\section{Conclusion}

In conclusion, we have measured the flow of dense colloidal suspensions, in microfluidic drums after an initial inclination angle, for different salt concentrations. The flowing curves show two regimes: a ``fast avalanche'' regime above a critical angle $\theta_c$, and a ``slow creep'' regime (logarithmic in time) below $\theta_c$. We observe that the flowing behavior is strongly modified by the ionic strength of the suspension. As the salt concentration increases, the initial time delay $\tau_S$ before the fast avalanche regime increases, the speed of this regime $\Delta \theta/\Delta t$ decreases, and the critical angle $\theta_c$ increases. All those observations are well explained by the fact that ions added in solution screen the repulsive double-layer electrostatic forces between the colloidal particles, which increases the effective microscopic friction between the particles $\mu_p$ and the initial packing fraction of the suspension $\Phi_0$ . We have independently verified with AFM measurements that the particles roughness $r$ is consistent with the critical salt concentration $C _c\sim \SI{2.3e-2}{\mole\per\liter}$ at which we observe a transition from ``very flowing curves'' to ``almost not flowing curves'' (with particles agglomerates). We have also verified with direct confocal microscopy observations that the initial packing fraction of the sedimented suspension increases from $\Phi_0 \sim \SI{51}{\percent}$ in deionized water to $\Phi_0 \sim \SI{61}{\percent}$ in solution with ionic strength $\SI{1e-2}{\mole\per\liter}$. This explains why increasing dilatancy effects are observed when more salt is added to the suspension.

Finally, even though all our measurements seem to indicate that the microscopic friction between the particles $\mu_p$ is increased by the salt concentration $C$, we cannot conclude on the physical origin of this effective friction increase. Indeed, this effective friction might come from direct contact friction (if the particles surfaces roughness touch each others), or from indirect hydrodynamic interactions (either long-range pore pressure effects, or short-range lubrication effects). Numerical simulations tend to show that contact friction dominates over long-range hydrodynamics at high volume fraction~\cite{Gallier2018} ($\Phi \geq \SI{40}{\percent}$), and over both long-range and short-range hydrodynamics at low viscous number~\cite{Chevremont2019} ($J \leq 10^{-1}$). However, only direct measurement of the normal and tangential forces between two colloidal particles (such as those obtained with quartz-tuning fork atomic force microscopy~\cite{Comtet2017, Chatte2018}, or lateral force microscopy~\cite{Jones2003,Fernandez2015}), in different ionic strength suspensions, would be able to experimentally confirm this result in our system.

\section*{Author Contributions}
A.B. designed the study, and built the horizontal video-microscopy apparatus. R.F. fabricated the microfluidic samples. M.L. and A.B. performed and analyzed the microfluid drums measurements. A.P. performed and analyzed the AFM measurements. A.B. performed and analyzed the confocal microscope measurements. A.B. curated the data and Pyhthon analysis scripts for the open data repository. All authors contributed to the writing of the manuscript.

\section*{Data availability}
\new{All data presented in this article, as well as the associated Python analysis scripts, are freely available on Zenodo repository: \href{https://doi.org/10.5281/zenodo.10203682}{10.5281/zenodo.10203682} and \href{https://doi.org/10.5281/zenodo.10671985}{10.5281/zenodo.10671985}.} Further requests should be addressed to Antoine Bérut.

\section*{Conflicts of interest}
There are no conflicts to declare.

\section*{Acknowledgements}
The authors acknowledge the support of the French Agence Nationale de la Recherche (ANR), under grant ANR-21-CE30-0005 (JCJC MicroGraM).\\
The authors would like to thank Gilles Simon for his help with building the horizontal video-microscopy apparatus, as well as for manufacturing some of the mechanical pieces used in the set-up ; Mathieu Leocmach for his help with the confocal microscopy measurements ;  Marie-Charlotte Audry, Christophe Ybert, Anne-Laure Biance, and Yoël Forterre for fruitful scientific discussions.

\bibliography{biblio}

\end{document}